\documentclass[prb,twocolumn,aps]{revtex4}
\usepackage{graphicx}
\usepackage{bm}
\usepackage{amssymb}
\usepackage{amsmath}

\input{tcilatex}

\begin{document}

\title{Mott Insulator - Superfluid Transitions in a Two Band Model at Finite
Temperature and Possible Application to Supersolid $^{4}$He}
\author{Huai-Bin Zhuang$^{1}$, Michael Ma$^{2}$, Xi Dai$^{1}$, Fu-Chun Zhang$%
^{1}$}
\date{\today }
\affiliation{$^{1}$Centre of Theoretical and Computational Physics, and Department of
Physics, The University of Hong Kong, Hong Kong\\
$^2$ Department of Physics, University of Cincinnati, Cincinnati, Ohio 45221}

\begin{abstract}
We study Mott insulator - superfluid transition in a two-band boson Hubbard
model, which can be mapped onto a spin-1/2 XY model with spins coupled to an
additional Ising degree of freedom. By using a modified mean field theory
that include the effects of phase fluctuations, we show that the transition
is first order at both zero and finite temperatures. On the Mott insulator
side, there may be reentrance in phase transition. These features are
consequences of the underlying transition between competing defect poor and
defect rich phases. The relevance of the model and our results to supersolid
$^{4}$He and cold bosonic atoms in optical lattices are discussed.
\end{abstract}

\maketitle

\section{INTRODUCTION}

The Mott insulator- Superfluid (MI-SF) transition is one of the striking
phenomena arising from the many body physics of correlated boson systems on
a lattice\cite{Fisher 1989}. The physics of this transition is believed to
be relevant to not only systems with bosons as \textquotedblright elementary
particles\textquotedblright , but also condensed matter systems whose low
energy physics is governed by bosonic degrees of freedom, for example, an
array of quantum Josephson junctions. Recently, there has been a renewed
interest in the MI-SF transition due to its relevance to two experimental
systems. The first is supersolidity of solid $^{4}$He\cite{Kim Chan
Nature,Kim Chan Science} if one views the normal solid to supersolid
transition as a Bose condensation transition that occurs due to
delocalization of the $^{4}$He atoms without melting the underlying
crystalline lattice. The second is ultracold bosonic atoms in an optical
lattice\cite{Bloch Nature}.

The commonly accepted paradigm for the MI-SF transition is based on the
one-band boson Hubbard model. In that model, the MI-SF transition is a
continuous one with the Mott-Hubbard gap and the condensate density
increasing continuously from $0$ on the MI and the SF sides of the
transition respectively. Recently, Dai, Ma, and Zhang (DMZ)\cite{DMZ}
proposed a two-band Hubbard hamiltonian as a model for solid $^{4}$He, and
showed using a single-site mean field theory that the MI-supersolid
transition can be first order. In this paper, we seek to investigate the
validity of this claim beyond mean field theory by including the effects of
phase fluctuations of the condensate. Specifically, we consider the
following spin $S=\frac{1}{2}$ XY model on a lattice, where the spins are
coupled to an additional Ising degree of freedom
\begin{equation*}
H=\Delta \sum_{i}\hat{n}_{i}-h\sum_{i}\hat{S}_{i}^{z}\hat{n}%
_{i}-J\sum_{\left\langle ij\right\rangle }\left( \hat{S}_{i}^{+}\hat{S}%
_{j}^{-}+\hat{S}_{i}^{-}\hat{S}_{j}^{+}\right) \hat{n}_{i}\hat{n}_{j}
\end{equation*}
Here the Ising variable $\hat{n}_{i}=0,1$ and $\hat{S}_{i}^{+}$ and $\hat{S}%
_{i}^{-}$ are the raising and lowering operators. As a model of magnetism,
this Hamiltonan would describe a XY spin system in a transverse magnetic
field $h$, with $n_{i}=0,1$ representing the absence (presence) of a spin on
site $i.$ In this case $\Delta $ plays the role of a chemical potential for
the spin. For example, $H$ can act as a model for an electronic insulator
with two on-site electronic configurations, one magnetic and one
non-magnetic separated by energy $\Delta ,$ or it can describe a binary
alloy of magnetic and non-magnetic ions. However, for this work, we are more
interested in $H$ as a model of boson insulator-superfluid transition. Below
we will discuss the relevance of this model to the boson two-band Hubbard
model and to supersolid $^{4}$He and trapped bosons in an optical lattice.

\bigskip

\subsection{Supersolid $^{4}$He}

The possibility of supersolidity in $^{4}$He, which refers to the
coexistence of crystalline ordering and superfluidity, was proposed some
years ago by Andreev and Lifshitz\cite{Andreev Lifshitz 1969}, Chester\cite%
{Chester 1970} , and Leggett\cite{Leggett 1970}. The interest in such an
unique state of matter is recently rekindled following the report by Kim and
Chan\cite{Kim Chan Nature, Kim Chan Science} of the observation of
non-classical rotational inertia (NCRI) in solid $^{4}$He confined inside
porous media and in bulk solid $^{4}$He at low temperature. \ NCRI has since
been confirmed by other groups\cite{Rittner Reppy 2006, Shirahama 2006,
Kubota 2007}, although it is still debatable as to whether this is a bulk
equilibrium effect, or a result of non-equilibrium defects\cite{Rittner
Reppy 2006, Burovsky 2005}. Thus far, experiments have failed to detect
direct superflow or \textquotedblright superconductivity\textquotedblright
\cite{Day Beamish 2006, Sasaki 2006}. Therefore, it remains controversial as
to whether $^{4}$He is a supersolid at low temperature. Nevertheless, these
recent experiments provide the impetus for better theoretical understanding
of the mechanism for bulk equilibrium supersolidity. In their seminal paper,
Andreev and Lifshitz\cite{Andreev Lifshitz 1969} proposed that since $^{4}$%
He is a quantum solid with large zero point motion, a finite density of
defects may be present even in the ground state. Amongst these defects, the
most promising candidate is zero point vacancies, whose motion they argued
will be wave-like at low temperature and so will\ Bose condense. The Bose
condensation of the vacancies can then lead to superfluidity without
destroying the underlying crystalline ordering. From the point of view of
the presence of off-diagonal long ranged order (ODLRO) in Jastrow
wavefunctions describing solids, Chester\cite{Chester 1970} also conjectured
zero point vacancies as the mechanism for supersolidity. If vacancies are
present, the solid will be incommensurate, with the number of He atoms
different from the number of lattice sites. Recently, Anderson, Brinkman and
Huse (ABH)\cite{Anderson Brinkman Huse 2005} pointed to the $T^{7}$
correction to the specific heat as evidence of incommensurability in the
ground state.

While appealing, the idea of zero point vacancies is subject to stringent
constraint from both experiments and several computational calculations,
showing vacancies as activated from the commensurate solid. The activation
energy is found to be about 10 $K$ from X-ray diffraction\cite{Simmons 1989}
and about $15$ $K$ in simulations\cite{Clark Ceperley 2006}. This is at odds
with the Andreev and Lishitz's proposal that the commensurate ground state
is unstable with respect to spontaneous generation of vacancies. Recently, \
Dai, Ma, and Zhang (DMZ) \cite{DMZ} proposed a possible solution to the
quandary by including a lower energy defect that they called the exciton,
which is an on-site bound state of a vacancy and an interstitial.
Physically, this defect corresponds to a broadening distortion of the local
wavefunction. For example, if the defect free solid state is given by the
(unsymmeterized) Hartree-Nosanow wavefunction\cite{Nosanow 1962}
\begin{equation*}
\Psi _{0}=\prod_{i}\phi _{a}(r_{i}-R_{i})
\end{equation*}%
where $\phi _{a}$ is a localized wavefunction, then the state with one
exciton defect on the site $j$ can be written as
\begin{equation*}
\Psi _{0}=\phi _{b}(r_{j}-R_{j})\prod_{i\neq j}\phi _{a}(r_{i}-R_{i})
\end{equation*}%
where $\phi _{b}$ is also localized, but less so than $\phi _{a}.$ The key
idea is that an atom in the state $\phi _{b}$ can tunnel more effectively
into a neighboring vacancy site (and vice versa) than one in the state $\phi
_{a}$ due to the wavefunction broadening. As a result, even though vacancies
are activated in a defect free background, they can be spontaneous generated
in an exciton background. At low temperature, they will then Bose condense
and the condensation energy may overcome the energy cost of creating exciton
defects. In their theory, the commensurate solid would then be a metastable
state, while the true ground state will be an incommensurate one with finite
densities of vacancies and excitons. The metastability of the commensurate
solid has also been proposed by Andeson, Brinkman, and Huse~\cite{Anderson
Brinkman Huse 2005}.

As a model for their theory, DMZ consider a two-band Hubbard model, in which
the defect free (DF) state is considered the \textquotedblright
vacuum\textquotedblright , and the boson operators are operators that create
and destroy defects. In this paper, it is more convenient to take the vacuum
as the physical vacuum for $^{4}$He atoms, in which case the DMZ Hamiltonian
becomes
\begin{eqnarray}
H &=&\sum_{i}\left( \varepsilon _{a}\hat{a}_{i}^{+}\hat{a}_{i}+\varepsilon
_{b}\hat{b}_{i}^{+}\hat{b}_{i}+U\hat{n}_{ai}\hat{n}_{bi}\right) \\
&&-\sum_{\left\langle ij\right\rangle }(t_{a}\hat{a}_{i}^{+}\hat{a}_{j}+t_{b}%
\hat{b}_{i}^{+}\hat{b}_{j}+h.c.)  \notag
\end{eqnarray}%
Here, the strong repulsion between $^{4}$He atoms in close proximity is
modeled by taking $\hat{a}$ and $\hat{b}$, field operators for $^{4}$He atom
in the $a$ and $b$ states, as hard-core boson operators. $U$ is the
repulsion between a $^{4}$He atom occupying $\phi _{a}$ and one occupying $%
\phi _{b}$ on the same site, and is of the order of the local interstitial
activation energy. Of the various energies in the problem, this is by far
the largest ($\sim 50$ $K$), and for the purpose of this work we will
consider $U\rightarrow \infty ,$ so each site can at most be singly occupied
by a $^{4}$He atom. In other words, we neglect the possibility of
interstitials. $t_{a}$ and $t_{b}$ denote the hopping amplitudes of the
tunnelling process between neighboring sites from $a$ to $a$-state and from $%
b$ to $b$-state respectively ($\left\langle ij\right\rangle $ indicates
nearest neighbors). The $b$-boson has higher on-site energy ($\varepsilon
_{b}>\varepsilon _{a}),$ but also a higher hopping amplitude ($t_{b}>t_{a}).$
The hard core nature of the bosons implies that hopping is only possible
between an occupied site and a neighboring vacancy.

In the torsional oscillator experiment setup used to observe NCRI, the
density of $^{4}$He atoms is held fixed. However, because the lattice
constant of the solid is determined by minimizing the free energy, the
number of bosons per site $\left\langle \hat{n}_{a}\right\rangle
+\left\langle \hat{n}_{b}\right\rangle $ is not externally imposed. Within
this model, if $\left\langle \hat{n}_{a}\right\rangle +\left\langle \hat{n}%
_{b}\right\rangle =1,$ the $^{4}$He atoms form a commensurate solid. If
further $\left\langle \hat{n}_{b}\right\rangle =0,$ the commensurate solid
is defect free (DF), while if $\left\langle \hat{n}_{a}\right\rangle =0,$ it
is an exciton solid. However, if $\left\langle \hat{n}_{a}\right\rangle
+\left\langle \hat{n}_{b}\right\rangle <1,$ then the solid is an
incommensurate one with a finite vacancy density. The excitation energy of a
single vacancy from the DF state is $\left\vert \varepsilon _{a}\right\vert
-zt_{a},$ where $z$ is the coordination number. The Andreev-Lifshitz vacancy
mechanism for supersolidity would require this to be negative, which
evidently is not the case. On the other hand, the instability criteria for
generating a vacancy above the exciton solid is $\left\vert \varepsilon
_{a}\right\vert -zt_{b},$ which is easier to satisfy. Vacancies can then
Bose condense above the exciton background and the condensation energy gain
may be sufficient to overcome the exciton energy $\Delta =\varepsilon
_{b}-\varepsilon _{a}.$ Should this be the case, the DF state will be
metastable. In their theory, DMZ identify the DF state as the normal solid
at $T=0$, and the exciton state with Bose condensation of vacancies as the
supersolid. Using a single-site mean field theory, DMZ showed that as
parameters in the mode are tuned (corresponding experimentally to for
example changing the pressure), there is a transition at $T=0$ from the DF
normal solid to the supersolid state described above. Since the DF state is
metastable, this transition is naturally first order, and involves a jump
not only in the Bose condensed amplitude, but also in the densities of both
vacancies and exciton defects. Hence the supersolid transition is
accompanied by a commensurate-incommensurate transition, as well as a change
in local density profile of the $^{4}$He atoms. Because the normal solid is
commensurate, it explains why experimental measurements performed on the
normal solid do not show an appreciable density of vacancies.

Because the results of DMZ are based on a single-site mean field theory, it
is of interest to examine if they hold up against the effects of both
quantum and thermal fluctuations. At low temperature, the dominating effects
should be due to phase fluctuations of the condensate, since they correspond
to gapless modes. Experimentally, at the lowest temperature, NCRI in solid $%
^{4}$He is observed even at the highest pressure achievable in the
laboratory. Thus, the $T=0$ first order supersolid-normal solid transition
predicted by DMZ cannot be tested at present. It is therefore important to
determine the nature of the supersolid-normal solid transition at finite
temperature within DMZ's theory to compare to experiments. We confirm that
when phase fluctuations are included, the finite temperature transition
remains first order. Moreover, we find that fluctuations give rise to a
'reentrance' phenomenon.

\bigskip

\subsection{Trapped Bosons in Optical Lattice}

For solid $^{4}$He, the lattice constant is self-adjusted to minimize the
free energy. We can also consider a system of bosons in the presence of an
external periodic potential, in which case the lattice constant will be
externally imposed. An interesting realization of such a system has been
achieved recently in ultracold trapped bosons in an optical lattice\cite%
{Jaksch 1998, Bloch Nature, Jaksch 2004}. By superimposing counter
propagating laser beams of wavelength $\lambda $ in different directions, an
effective periodic potential in one, two, or three dimensions can be
produced. In $3D,$ the potential has the form
\begin{equation*}
V(x,y,z)=V_{0x}\cos ^{2}(kx)+V_{0y}\cos ^{2}(ky)+V_{0z}\cos ^{2}(kz)
\end{equation*}%
where $k=2\pi /\lambda $ is the wavenumber, and the strengths of the
potential in the three directions are proportional to the laser intensity in
each direction and can therfore be tuned. For a single particle, this
periodic potential gives rise to the usual energy bands in the first
Brillouin zone. For $V_{0}$ large compared to the so-called recoil energy $%
E_{R}=\hslash ^{2}k^{2}/2m,$ the lower bands can be viewed as tight-binding
bands arising from a periodic array of harmonic wells. For deep potential of
cubic symmetry ($V_{0x}=V_{0y}=V_{0z}=V_{0}),$ the Wannier basis for these
bands can to a good approximation be constructed from the eigenstates of the
spherically symmetric harmonic oscillator potential $%
V_{har}(r)=V_{0}k^{2}r^{2}.$ The bands can then be denoted by the quantum
numbers $(n_{x},n_{y},n_{z})$ of these eignestates, and their eigenenergies $%
\left( n_{x}+n_{y}+n_{z}+1/2\right) \hslash \omega _{0},$ where $\hslash
\omega _{0}=\sqrt{4V_{0}E_{R}},$ provide approximate values for the band
gaps. It is convenient to consider the optical lattice experiments as being
performed at fixed chemical potential. \ In the case where $\hslash \omega
_{0}$ is much larger than other relevant energy scales, all the particles
will occupy the lowest band, and the system can be modeled by a single-band
boson Hubbard model\cite{Jaksch 2004}:
\begin{eqnarray*}
H &=&\sum_{i}\varepsilon _{a}\hat{a}_{i}^{+}\hat{a}_{i}+U\sum_{i}\hat{n}%
_{ai}(\hat{n}_{ai}-1) \\
&&-\sum_{\left\langle ij\right\rangle }(t_{a}\hat{a}_{i}^{+}\hat{a}%
_{j}+h.c.)-\mu \sum_{i}\left( \hat{a}_{i}^{+}\hat{a}_{i}\right)
\end{eqnarray*}%
with $\mu $ as the chemical potential. This is a well-studied model, and it
has been established that with increasing $t_{a}/U,$ there is a quantum
phase transition from a Mott insulator state with integer filling per site
to a superfluid (SF) state. The SF state may have commensurate or
incommensurate filling depending on $\mu $. Trapped bosons in the optical
lattice provides an elegant realization of this model which allows the
transition to be studied systematically in experiments.

Alternatively, we can consider the limit where the Hubbard $U$ is the
dominating energy. This can be achieved experimentally by tuning close to
the Feshbach resonance. In this case the\ Mott-Hubbard gap prevents any
double or higher occupancy $n_{i}\leq 1,$ and the bosons are hard-core. If
we restrict to the lowest band still, then the ground state is an insulator
for $\mu <\varepsilon _{a}-zt_{a}$ (vacuum) and $\mu >\varepsilon
_{a}+zt_{a} $ (one boson on each site), while if $\mu $ lies between these
two limits, the system is a superfluid with $\left\langle n_{i}\right\rangle
$ changing continuously from $0$ to $1$ as $\mu $ increases. Unlike the
MI-SF transition with changing $t_{a}/U,$ these insulator-SF transitions
caused by changing $\mu $ in the hard-core model are just by-products of
density transitions.

For $\mu >\varepsilon _{a}+zt_{a},$ one has essentially a filled band
insulator if we restrict to the lowest band. However, if the higher bands
are considered, then it is possible for bosons to delocalize and form a SF.
We emphasize from the start that this is not simply due to partially
occupying the higher bands while keeping the lowest band filled or even
transferring some bosons from the lowest band to a higher band, as the hard
core condition would still forbid any boson motion if every site is singly
occupied independent of which band it is in. Instead the boson
delocalization can only occur by introducing vacancies.

\begin{figure}[h]
\includegraphics[width=8.0cm,height=6.0cm]{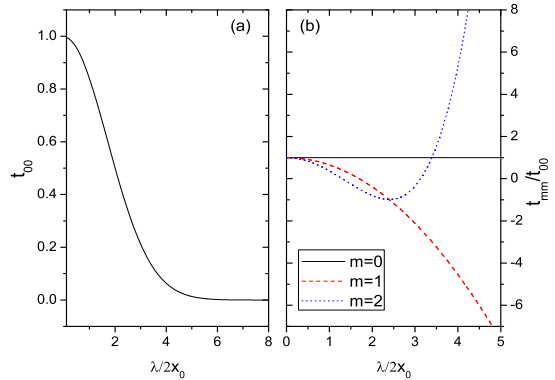}
\caption{(a) The overlap $t_{00}$ of the lowest energy states and (b) $%
t_{mm} $ $\left( m=0,1,2\right) $ scaled by $t_{00},$ as functions of the
distance $\protect\lambda /2$ between two neighboring centers in the unit of
Gaussian half-width $x_{0}$ for a 1D harmonic oscillator in estimating the
tunnelling amplitudes of bosons. We use $t_{mn}\left( a\right) =\protect\int%
_{-\infty }^{\infty }\protect\varphi _{m}^{\ast }\left( x\right) \protect%
\varphi _{n}\left( x-a\right) dx,$ where $\protect\varphi _{n}\left(
x-a\right) $ denotes the eigenstate of the n$^{th}$ energy level of the
harmonic oscillator with center $x=a.$ }
\label{1Dharmonic}
\end{figure}

We consider the case where in addition to the lowest band, we include one
higher band, with both strong enough intraband and interband on-site
repulsion between bosons to make them hard-core. One thus arrives at a
two-band boson model
\begin{eqnarray*}
H &=&\sum_{i}\left( \varepsilon _{a}\hat{a}_{i}^{+}\hat{a}_{i}+\varepsilon
_{b}\hat{b}_{i}^{+}\hat{b}_{i}\right) -\sum_{\left\langle ij\right\rangle
}(t_{a}\hat{a}_{i}^{+}\hat{a}_{j}+t_{b}\hat{b}_{i}^{+}\hat{b}_{j}+h.c.) \\
&&-\mu \sum_{i}\left( \hat{a}_{i}^{+}\hat{a}_{i}+\hat{b}_{i}^{+}\hat{b}%
_{i}\right)
\end{eqnarray*}%
with the constraint $n_{ai}+n_{bi}\leq 1.$ This Hamiltonian is of the same
form as the DMZ 2-band model for superolid except for the presence of the
chemical potential term, which can simply be absorbed into a redefinition of
$\varepsilon _{a}$ and $\varepsilon _{b}$ . The $\hat{a}$ and $\hat{b}$
operators are for the lowest and the higher bands respectively. In analogy
to the supersolid model, we will refer to a boson in the $b$-band as an
exciton. Because the Wannier state for the higher band is less localized,
the hopping amplitude $t_{b}$ will have a bigger magnitude than $t_{a}$ if
the lattice constant $\lambda /2$ is sufficiently large. The ratio of the
hopping amplitudes can be estimated by looking at the ratio of overlaps
between states on neighboring sites. Using this estimate, we calculate the
quantity $t_{mn}/t_{00},$ where $t_{mn}$ is the probability density of the
boson tunneling from eigenstate $n$ of a well into eigenstate $m$ of a
neighboring well. In Fig. \ref{1Dharmonic}(a), we plot $t_{00}$ as a
function of the Gaussian half-width of the harmonic oscillator. The
dependence of the ratio $t_{mm}/t_{{00}}$ on the distance between wells, $%
\lambda /2,$ is shown in Fig. \ref{1Dharmonic}(b) for $m=0,1,2$ in one
dimension. If we identify $a$ as the $n=0$ state, and $b$ as either the $n=1$
or $n=2$ state, then we can see clearly that $t_{b}$ can become
significantly larger than $t_{a}$ as $\lambda /2x_{0}$ increases, where $%
x_{0}$ is the Gaussian half-width of the $n=0$ harmonic oscillator
eigenstate.

Instead of the lowest and a higher band of a single type of bosons, the
two-band model also acts as the model for the case of two different types of
bosons ($a,b)$ each restricted to its respective lowest band in the optical
lattice. For example, we may have bosons of two different masses ($%
m_{a}>m_{b})$. If we assume the laser beams produce the same periodic
potential on them, then we have $\Delta =\varepsilon _{b}-\varepsilon
_{a}=\hslash \omega _{0b}-$ $\hslash \omega _{0a},$ where $\hslash \omega
_{0a,b}\varpropto \sqrt{\frac{1}{m_{a,b}}}$ are the harmonic oscillator
frequencies for mass $m_{a}$ and $m_{b}$ respectively. We also have $%
t_{b}>t_{a},$ due to larger zero point motion of the $b-$boson. \ However,
because the tunneling amplitude $t_{00}^{a,b}$ is exponentially sensitive to
the gaussian half-width of the harmonic oscillator wavefunction, we expect
the difference in tunneling amplitudes to be the more significant effect.

Before we discuss our calculation for the two-band model, we should clarify
the terminology used below. In both solid $^{4}$He and trapped bosons in
optical lattice, Galilean invariance is violated. In the optical lattice
case, the lack of translational invariance is externally imposed, so when
bosons Bose condense, we consider it to be a superfluid, not a supersolid.
In the case of solid $^{4}$He, the translational invariance is spontaneously
broken by the solid, and if Bose condensation occurs without the solid
melting, we have a supersolid, to be distinguished from the superfluid,
which refers to the situation in the liquid. However, since we are not
interested in the melting transition, and since the two-band model is
applied to both solid $^{4}$He and optical lattice, we will simply refer to
the Bose condensed phase as the SF phase in both systems for convenience in
what follows.

\bigskip

\section{Modified Mean Field Theory}

Our primary goal is to analyze the above 2-band boson Hubbard model to
include phase fluctuations. Since we do not expect the precise details of
the individual parameters in the Hamiltonian other than the gross features
described above to be crucial, we take $t_{a}=0$ for convenience. \ The
boson model can then be mapped onto a spin $S=1/2$ XY model coupled to an
Ising degree of freedom as follows. \ \ \
\begin{eqnarray}
H &=&\sum_{i}\left( \varepsilon _{a}-\mu \right) +\sum_{i}\left( \Delta +%
\frac{h}{2}\right) \hat{n}_{i}-h\sum_{i}\hat{S}_{iz}\hat{n}_{i}
\label{spinxy} \\
&&-t_{b}\sum_{\left\langle ij\right\rangle }\left( \hat{S}_{i}^{+}\hat{S}%
_{j}^{-}+\hat{S}_{i}^{-}\hat{S}_{j}^{+}\right) \hat{n}_{i}\hat{n}_{j},
\notag
\end{eqnarray}%
with $h=\mu -(\varepsilon _{a}+\Delta ).$ The mapping between the original
boson operators and the new set of operators $\hat{n}$ and $\hat{S}$ is
\begin{equation}
\hat{n}_{i}\,\left\vert 0\right\rangle =\left\vert 0\right\rangle ,\ \ \ \,%
\hat{n}_{i}\hat{b}_{i}^{+}\left\vert 0\right\rangle =\hat{b}%
_{i}^{+}\left\vert 0\right\rangle ,\,\ \ \,\hat{n}_{i}\hat{a}%
_{i}^{+}\left\vert 0\right\rangle =0
\end{equation}%
\begin{equation}
\hat{S}_{i}^{z}=\hat{b}_{i}^{+}\hat{b}_{i}-\frac{1}{2},\text{ \ }\hat{S}%
_{i}^{+}=\hat{b}_{i}^{+},\text{ \ }\hat{S}_{i}^{-}=\hat{b}_{i}
\end{equation}%
where $\left\vert 0\right\rangle $ is the empty site, i.e. vacancy state.
The Ising operator $\hat{n}_{i}$ has eigenvalues $0$ and $1.$ For bosons, it
acts as the defect (vacancy or exciton) occupation number operator. The spin
$1/2$ operator $\hat{S}_{i}$ acts on the Hilbert space of the $\left\vert
0\right\rangle $ (vacancy) and $\hat{b}_{i}^{+}\left\vert 0\right\rangle $
(exciton) doublet, which is assumed to be closer to each other in energy
than to that of $\hat{a}_{i}^{+}\left\vert 0\right\rangle .$

\bigskip Because we take $t_{a}=0,$ the $a$-bosons cannot Bose condense, and
the Bose condensation order parameter, taken to be real, is given by $%
\left\langle \hat{b}_{i}\right\rangle =\left\langle \hat{b}%
_{i}^{+}\right\rangle =\left\langle \hat{S}_{ix}\right\rangle .$ In this
representation, the single site MFT of DMZ would correspond to approximating
$\hat{S}_{i}^{+}\hat{S}_{j}^{-}\hat{n}_{i}\hat{n}_{j}$ by
\begin{eqnarray*}
\hat{S}_{i}^{+}\hat{S}_{j}^{-}\hat{n}_{i}\hat{n}_{j} &\approx &\hat{S}%
_{i}^{+}\left\langle \hat{S}_{j}^{-}\right\rangle \left\langle \hat{n}%
_{i}\right\rangle \left\langle \hat{n}_{j}\right\rangle +\hat{S}%
_{j}^{-}\left\langle \hat{S}_{i}^{+}\right\rangle \left\langle \hat{n}%
_{i}\right\rangle \left\langle \hat{n}_{j}\right\rangle \\
&&+\hat{n}_{i}\left\langle \hat{S}_{i}^{+}\right\rangle \left\langle \hat{S}%
_{j}^{-}\right\rangle \left\langle \hat{n}_{j}\right\rangle +\hat{n}%
_{j}\left\langle \hat{S}_{i}^{+}\right\rangle \left\langle \hat{S}%
_{j}^{-}\right\rangle \left\langle \hat{n}_{i}\right\rangle \\
&&-3\left\langle \hat{S}_{i}^{+}\right\rangle \left\langle \hat{S}%
_{j}^{-}\right\rangle \left\langle \hat{n}_{i}\right\rangle \left\langle
\hat{n}_{j}\right\rangle
\end{eqnarray*}%
If we apply the single-site mean field theory to this model, we simply
recover the results of DMZ for the choice of parameters used here. Such MFT
of course neglects fluctuations completely, and it is legitimate to question
if the results remain valid when fluctuations are included. At low
temperature, the dominating fluctuations in three dimensions should be those
from gapless excitations, which are the phase modes of the Bose condensate,
or spin waves in the spin language. In order to include the effects of phase
fluctusations, we modify the mean field theory as follows
\begin{eqnarray}
\hat{S}_{i}\hat{S}_{j}\hat{n}_{i}\hat{n}_{j} &=&\left\langle \hat{S}_{i}\hat{%
S}_{j}\right\rangle \left( \left\langle \hat{n}_{i}\right\rangle \hat{n}_{j}+%
\hat{n}_{i}\left\langle \hat{n}_{j}\right\rangle \right) \\
&&+\hat{S}_{i}\hat{S}_{j}\left\langle \hat{n}_{i}\right\rangle \left\langle
\hat{n}_{j}\right\rangle -2\left\langle \hat{S}_{i}\hat{S}_{j}\right\rangle
\left\langle \hat{n}_{i}\right\rangle \left\langle \hat{n}_{j}\right\rangle
\notag
\end{eqnarray}%
\ That is, correlations between the Ising degrees of freedom on different
sites and between the Ising and spin degrees of freedom on same or different
sites are still ignored. However, correlations between the spin degrees of
freedom on different sites will be included to allow for spin wave physics.
The modified mean field Hamiltonian can now be written as
\begin{equation*}
H=H_{n}+H_{s}+C.
\end{equation*}%
where
\begin{equation}
H_{s}=-h\bar{n}\sum_{i}\hat{S}_{i}^{z}-J\sum_{\left\langle ij\right\rangle
}\left( \hat{S}_{i}^{x}\hat{S}_{j}^{x}+\hat{S}_{i}^{y}\hat{S}_{j}^{y}\right)
\end{equation}%
\begin{equation}
H_{n}=\left( \Delta +\frac{h}{2}-hM-2t_{b}Z\bar{n}B\right) \sum_{i}\hat{n}%
_{i}
\end{equation}%
\begin{equation}
C=\sum_{i}\{\left( \varepsilon _{a}-\mu \right) +h\bar{n}M+2t_{b}Z\overline{n%
}^{2}B\}
\end{equation}%
where $J=2t_{b}\bar{n}^{2}$ and $Z$ is the co-ordination number.
\begin{eqnarray*}
M &=&\left\langle \hat{S}_{i}^{z}\right\rangle \\
B &=&\left\langle \hat{S}_{i}^{x}\hat{S}_{i+\delta }^{x}\right\rangle
+\left\langle \hat{S}_{i}^{y}\hat{S}_{i+\delta }^{y}\right\rangle \\
\overline{n} &=&\left\langle \hat{n}_{i}\right\rangle .
\end{eqnarray*}%
are parameters independent of site number and to be determined
self-consistently. \ It can be seen that the spin part Hamiltonian $H_{s}$
is a ferromagnetic XY model in a transverse field with exchange coupling $J$
.

$H_{n}$ is a single-site Hamiltonian and $\left\langle \hat{n}%
_{i}\right\rangle $ is easily calculated. The self-consistent equation for $%
\overline{n}$ is then given by
\begin{equation}
\overline{n}=\frac{Tr\left[ n_{i}e^{-\beta H}\right] }{Tr\left[ e^{-\beta H}%
\right] }=\frac{\exp \left( -\beta E_{1}\right) }{\frac{1}{2}+\exp \left(
-\beta E_{1}\right) }  \label{nEqu}
\end{equation}
where
\begin{equation*}
E_{1}=\Delta +\frac{h}{2}-hM-2t_{b}Z\bar{n}B
\end{equation*}
The factor $\frac{1}{2}$ in the denominator arises from the $n_{i}=0$ state
being a singlet state while the $n_{i}=1$ state is a doublet.

The most important feature of $H_{s}$ is its continuous XY symmetry which
gives rise to gapless excitations (Goldstone modes) in the ordered state.
This physics will not be affected by the transverse field $h$ provided $%
\left| h\right| $ is not too big. Thus, as further simplification, we
consider in this work the zero field case, or in other words the case where
the $b$-state and the vacancy are degenerate. \ \ Setting $h=0,$ the three
parts of the MF Hamiltonian becomes
\begin{equation}
H_{s}=-J\sum_{\left\langle ij\right\rangle }\left( \hat{S}_{i}^{x}\hat{S}%
_{j}^{x}+\hat{S}_{i}^{y}\hat{S}_{j}^{y}\right)
\end{equation}
\begin{equation}
H_{n}=\left( \Delta -2t_{b}Z\bar{n}B\right) \sum_{i}\hat{n}_{i}
\end{equation}
\begin{equation}
C=N\left[ \left( \varepsilon _{a}-\mu \right) +2t_{b}Z\bar{n}^{2}B\right]
\end{equation}

\

\subsection{The Modified Spin-Wave Method}

After the modified mean field decoupling, $H_{s}$ remains a many body
Hamiltonian. Since our main goal is to include the effects of the condensate
phase fluctuations, which in the spin language implies fluctuations of spin
waves, it is natural to use the spin wave approximation. The spin wave
theory is known to be reliable in $3D,$ where long range order persists at
low temperature. However, it is seen that the coupling $J$ is proportional
to $\bar{n}^{2},$ which is itself temperature dependent, and as we will see,
will imply that we need to address $H_{s}$ both below and above the spin
ordering temperature. One possible method that can be used is the Schwinger
boson MFT, but it has been shown that this method is problematic for
temperature comparable or larger than $J$. \cite{theja} Instead, we use a
modified spin wave method similar to the approaches introduced by Takahasi%
\cite{Takahashi 1987,Takahashi 1989} and by Tang and Hirsch\cite{Hirsch Tang
1989}.

The calculation will be performed on the 3D cubic lattice. For the case of
bosons on the optical lattice, this is the actual lattice that has been
studied experimentally. For solid $^{4}$He, the lattice is hcp, but since $%
H_{s}$ is ferromagnetic, the physics does not depend on the precise lattice
structure beyond the co-ordination number $Z.$ To perform the spin wave
theory, it is convenient\cite{Gomez-Santos Joannopoulos 1987} to globally
rotate the spins about the spin y-axis and rewrite $H_{s}$ as
\begin{equation*}
H_{s}=-J\sum_{l\delta }\left( \hat{P}_{l}^{z}\hat{P}_{l+\delta }^{z}+\hat{P}%
_{l}^{x}\hat{P}_{l+\delta }^{x}\right)
\end{equation*}
where the $\hat{P}$'s are spin operators in the rotated frame. The classical
ground state can then be taken as having all the spins pointing in the $+z$
direction. Next, the Holstein-Primakoff transformation is defined
\begin{eqnarray}
P_{l}^{+} &=&\sqrt{2S-c_{l}^{+}c_{l}}c_{l}  \notag \\
P_{l}^{-} &=&c_{l}^{+}\sqrt{2S-c_{l}^{+}c_{l}} \\
P_{l}^{z} &=&S-c_{l}^{+}c_{l}  \notag
\end{eqnarray}
The linearized spin wave Hamiltonian $H_{SW}$ is obtained by expanding the
square roots and keeping to quadratic order in the bosonic $c$ operators.
This quadratic Hamiltonian can then be solved using Bogliubov transformation
\begin{equation*}
d_{k}=\cosh \theta _{k}c_{k}-\sinh \theta _{k}c_{-k}^{+},\text{ \ }%
d_{k}^{+}=\cosh \theta _{k}c_{k}^{+}-\sinh \theta _{k}c_{-k}\text{\ }
\end{equation*}
where
\begin{equation*}
c_{l}=\frac{1}{\sqrt{N}}\sum_{k}e^{ik\cdot l}c_{k},\text{ \ }c_{l}^{+}=\frac{
1}{\sqrt{N}}\sum_{k}e^{-ik\cdot l}c_{k}^{+}
\end{equation*}

Unlike the Heisenberg ferromagnet, where the classical ground state is also
the quantum mechanical one, quantum fluctuations are present for the XY
ferromagnet. As a result $\left\langle P_{l}^{z}\right\rangle <1/2$ even for
the ground state. Provided the difference is small though, the classical
ground state can be considered to be a good approximation and the spin wave
theory should be reliable. As temperature increases, however, the amount of
fluctuations increase, and eventually $\left\langle P_{l}^{z}\right\rangle $
calculated by spin wave theory becomes $<0,$ which clearly indicates spin
wave theory ceases to be valid. However, spin wave theory can be applied if
a Lagrangian multiplier or magnon chemical potential $\lambda $ is added to
restrict the total number of magnons excited. \ The modified spin wave
Hamiltonian is given by

\begin{equation}
H_{SW}^{\prime }=H_{SW}-\lambda \sum_{l}\left( S-c_{l}^{+}c_{l}\right)
\end{equation}%
with the provision that if $\left\langle c_{l}^{+}c_{l}\right\rangle \leq
1/2 $ when calculated without $\lambda ,$ then $\lambda =0,$ and if $%
\left\langle c_{l}^{+}c_{l}\right\rangle >1/2,$ then $\lambda $ becomes
non-zero and is adjusted to enforce $\left\langle
c_{l}^{+}c_{l}\right\rangle =1/2,$ i.e. $\left\langle P_{l}^{z}\right\rangle
=0.$ After Bogliubov transformation, $\;H_{SW}^{\prime }$ becomes
diagonalized
\begin{equation}
H_{SW}^{\prime }=-NZ\frac{3J}{8}-\lambda N+\sum_{k}E_{k}\left(
d_{k}^{+}d_{k}+\frac{1}{2}\right)
\end{equation}%
where the magnon excitation energy
\begin{equation}
E_{k}=\frac{J}{2}Z\left( 1+\lambda ^{\prime }\right) \sqrt{1-\frac{\gamma
_{k}}{\left( 1+\lambda ^{\prime }\right) }}
\end{equation}%
and $\lambda ^{\prime }=\frac{2\lambda }{J},$ $\gamma _{k}=\frac{1}{Z}\sum
e^{ik\cdot \delta }.$ \ The Bogliubov coefficients are found to be
\begin{equation}
\cosh 2\theta _{k}=\frac{\alpha _{k}}{E_{k}},\sinh 2\theta _{k}=\frac{2\beta
_{k}}{E_{k}}
\end{equation}%
with
\begin{equation}
\alpha _{k}=t_{b}Z\bar{n}^{2}-\frac{1}{2}t_{b}Z\bar{n}^{2}\gamma
_{k}+\lambda ,\text{ \ }\beta _{k}=\frac{1}{4}t_{b}Z\bar{n}^{2}\gamma _{k}
\end{equation}

\begin{figure}[tbp]
\includegraphics[width=7.0cm,height=8.0cm]{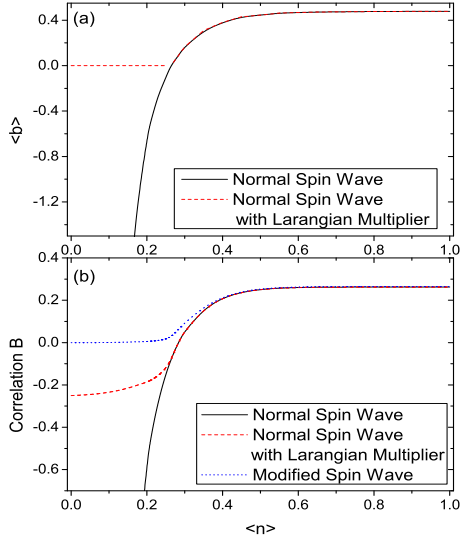}
\caption{The modifications made in solving the spin XY model with
'vacancies'. \ (a) The superfluid order parameter $<b>$ and (b) the spin
correlation $B=\left\langle
S_{i}^{x}S_{j}^{x}+S_{i}^{y}S_{j}^{y}\right\rangle $ before and after
modifcations as functions of $\bar{n}$ at $kT=0.05t_{b}Z,$ where $\bar{n}$
is not yet self-consistent solutions. \ $<b>$ is tuned to be non-negative at
small $\bar{n}$'s by a Lagrangian multiplier. \ The correlation $B$ is
further corrected by the modified spin wave method. \ }
\label{modified spin wave}
\end{figure}

The Bose condensed order parameter is \
\begin{equation}
\left\langle \widehat{b}\right\rangle =\left\langle P^{z}\right\rangle =1-%
\frac{1}{N}\sum_{k}\cosh 2\theta _{k}\left( n_{k}+\frac{1}{2}\right)
\end{equation}
where
\begin{equation}
n_{k}=\left\langle d_{k}^{+}d_{k}\right\rangle =\left( \exp \left( \beta
E_{k}\right) -1\right) ^{-1}
\end{equation}
is the Bose-Einstein distribution of the magnons, $\beta ^{-1}=T$ (we pick
energy unit so that the Boltzmann constant $k_{B}=1)$. \ The dependence of $%
\left\langle \widehat{b}\right\rangle $ on $\overline{n}$ as calculated from
linearized spin wave theory with and without the Lagrange multiplier at a
fixe temperature is shown in Fig. \ref{modified spin wave}a. When the order
parameter $\left\langle \widehat{b}\right\rangle $ is positive, denoting the
SF phase, $\lambda =0$ and the dispersion $E_{k}$ represents a gapless mode
and is linear in $k$ at small wave number $k$. \ On the contrary, the
excitation energy $E_{k}$ has a finite gap for nonzero $\lambda $ and hence
for zero $\left\langle \widehat{b}\right\rangle ,$ indicating a MI phase.

\bigskip Before we set off to calculate the self-consistent parameters $B$
and $\bar{n}$, a glance at the bosonic distribution $n_{k}$ alerts us that\
the spin wave method with chemical potential is still problematic. \ Since $%
J\varpropto \bar{n}^{2},$ the effective spin temperature is $T^{\prime }=T/%
\bar{n}^{2}.$ \ Though the Lagrangian multiplier prescription is reliable up
to $T\sim J=2t_{b}\bar{n}^{2},$ it fails when $T>>J.$ However, the necessity
to consider small $\bar{n}$ values can mean $T$ $>>J$ even at low
temperature. One major problem with the modified spin wave method discussed
so far is that at high $T^{\prime },$ the nearest neighbor spin-spin
correlation function $B$ can actually take on the wrong sign. This problem
has been recognized in literature, and a remedy has been proposed\cite%
{Takahashi 1987, Takahashi 1989}. \ \ The correct sign of $B$ can be
maintianed if when calculating the spin correlation\ $\left\langle
P_{l}^{z}P_{l+\delta }^{z}+P_{l}^{x}P_{l+\delta }^{x}\right\rangle $ , one
keeps not just the quadratic, but the quartic order in HP boson operators:
\begin{eqnarray}
&&\left\langle P_{l}^{z}P_{l+\delta }^{z}+P_{l}^{x}P_{l+\delta
}^{x}\right\rangle \\
&=&P^{2}-P[\left\langle c_{i}^{+}c_{i}\right\rangle +\left\langle
c_{j}^{+}c_{j}\right\rangle  \notag \\
&&-\frac{1}{2}\left( \left\langle c_{i}c_{j}\right\rangle +\left\langle
c_{i}^{+}c_{j}^{+}\right\rangle +\left\langle c_{i}c_{j}^{+}\right\rangle
+\left\langle c_{i}^{+}c_{j}\right\rangle \right) ]  \notag \\
&&+\left\langle c_{i}^{+}c_{i}c_{j}^{+}c_{j}\right\rangle -\frac{1}{8}%
[\left\langle c_{i}^{+}c_{i}c_{i}c_{j}\right\rangle +\left\langle
c_{i}c_{j}^{+}c_{j}c_{j}\right\rangle  \notag \\
&&+\left\langle c_{i}^{+}c_{i}^{+}c_{i}c_{j}^{+}\right\rangle +\left\langle
c_{i}^{+}c_{j}^{+}c_{j}^{+}c_{j}\right\rangle +\left\langle
c_{i}^{+}c_{i}c_{i}c_{j}^{+}\right\rangle  \notag \\
&&+\left\langle c_{i}c_{j}^{+}c_{j}^{+}c_{j}\right\rangle +\left\langle
c_{i}^{+}c_{i}^{+}c_{i}c_{j}\right\rangle +\left\langle
c_{i}^{+}c_{j}^{+}c_{j}c_{j}\right\rangle ]+O\left( P^{-1}\right)  \notag
\end{eqnarray}%
The quartic terms of $c$ operators are then evaluated by Hartree-Fock (HF)
decoupling
\begin{equation}
\left\langle c_{i}^{+}c_{i}c_{j}^{+}c_{j}\right\rangle \approx \left\langle
c_{i}^{+}c_{i}\right\rangle \left\langle c_{j}^{+}c_{j}\right\rangle
+\left\langle c_{i}^{+}c_{j}\right\rangle \left\langle
c_{i}c_{j}^{+}\right\rangle +\left\langle c_{i}^{+}c_{j}^{+}\right\rangle
\left\langle c_{i}c_{j}\right\rangle
\end{equation}

In the theory by Takahasi, these HF terms need to be determined
self-consistently, in essence renormalizing the quadratic Hamiltonian $%
H_{SW}^{\prime }.$ That is too complicated to do in our problem, so we just
evaluate them using the unrenormalized $H_{SW}^{\prime },$ in line with the
scheme proposed by Hirsch and Tang\cite{Hirsch Tang 1989}$.$ \ Using $%
H_{SW}^{\prime }$, the HF terms can be derived as
\begin{eqnarray}
\left\langle c_{i}^{+}c_{j}\right\rangle &=&f\left( r_{i}-r_{j}\right) -%
\frac{1}{2}\delta _{ij} \\
\left\langle c_{i}c_{j}^{+}\right\rangle &=&f\left( r_{i}-r_{j}\right) +%
\frac{1}{2}\delta _{ij}  \notag \\
\left\langle c_{i}c_{j}\right\rangle &=&\left\langle
c_{i}^{+}c_{j}^{+}\right\rangle =g\left( r_{i}-r_{j}\right)  \notag
\end{eqnarray}%
where
\begin{eqnarray}
f\left( r_{i}-r_{j}\right) &=&\frac{1}{N}\sum_{k}\exp \left( ik\cdot \left(
r_{i}-r_{j}\right) \right) \cosh 2\theta _{k}  \notag \\
&&\times \left( \left\langle d_{k}^{+}d_{k}\right\rangle +\frac{1}{2}\right)
\notag \\
g\left( r_{i}-r_{j}\right) &=&\frac{1}{N}\sum_{k}\exp \left( ik\cdot \left(
r_{i}-r_{j}\right) \right) \sinh 2\theta _{k}  \notag \\
&&\times \left( \left\langle d_{k}^{+}d_{k}\right\rangle +\frac{1}{2}\right)
\end{eqnarray}%
Consequently, the spin correlation becomes
\begin{eqnarray}
&&B\left( \bar{n},T\right) \\
&=&\left\langle P_{l}^{z}P_{l+\delta }^{z}+P_{l}^{x}P_{l+\delta
}^{x}\right\rangle  \notag \\
&=&\left( S-\left( f\left( 0\right) -\frac{1}{2}-\frac{1}{2}g\left( \delta
\right) -\frac{1}{2}f\left( \delta \right) \right) \right) ^{2}  \notag \\
&&+\frac{1}{4}\left( g\left( \delta \right) -f\left( \delta \right) \right)
^{2}+\frac{1}{2}\left( g\left( \delta \right) -\frac{1}{2}g\left( 0\right)
\right) ^{2}  \notag \\
&&+\frac{1}{2}\left( f\left( \delta \right) -\frac{1}{2}g\left( 0\right)
\right) ^{2}-\frac{1}{4}g\left( 0\right) ^{2}  \notag
\end{eqnarray}%
Though $B$ does not appear in the form of a complete square as the case of
Heisenberg ferromagnet\cite{Takahashi 1987}, the inclusion of the quartic
terms nevertheless ensures that it has the right sign at all effective spin
temperature $T^{\prime }$ (See Fig. \ref{modified spin wave}b). Together
with the equation for $\overline{n}$\ Eq. (\ref{nEqu}), this equation for $B$
provide the self-equations for $\overline{n}$\ and $B.$

Finally, within our MFT, the free energy per site is
\begin{eqnarray}
G\left( T\right) &=&-\frac{kT}{N}\ln Tr\exp \left[ -\beta \left(
H_{n}+C+H_{s}\right) \right]  \label{free energy} \\
&=&-kT\ln \left( \frac{1}{2}+e^{-\beta E_{1}}\right) +t_{b}Z\bar{n}%
^{2}B\left( \bar{n},T\right) -TS_{s}  \notag
\end{eqnarray}
where $S_{s}$ is the spin entropy. The free energy is useful in case where
there is more than one MF solutions to determine which is the most stable
solution. Within our MF approach this is derived in principle from $%
H_{SW}^{\prime },$ which gives another problem. \ At high temperature, the
spin entropy for a spin $1/2$ system saturates to $\ln 2.$ However, using $%
H_{SW}^{\prime }$, the spin entropy is $S_{s}=\sum_{k}\{\left(
1+n_{k}\right) \ln \left( 1+n_{k}\right) -n_{k}\ln n_{k}\}$ and can exceed
this saturated value due to large fluctuations even when the Legrange
multiplier limits the average value of $n_{k}.$ Since the effective spin
temperature $T^{\prime }$ can be enormous at small $\bar{n},$ this problem
is relevant when we need to compare the free energy of the $\bar{n}\approx 0$
to the $\bar{n}\approx 1$ MF solutions at temperature of interest. Our
remedy for this problem is to impose an upper limit $S_{s}=k_{B}\ln 2$ when
the calculated value of $S_{s}$ exceeds that value.

\section{Results and Discussions}

The results obtained for the 3D SC lattice are as follows. The main result
will be that i) the $T=0$ first order transition obtained by DMZ in their
single-site MFT remains robust with phase fluctuations; ii) close to the $%
T=0 $ transition point, the finite temperature transition will also be first
order\cite{footnote}, and iii) the stablization of the ordered phase gives
rise to reentrance with increasing temperature into the SF phase on the MI
side of the $T=0$ transition. We first present our results for bosons in
optical lattice in which case the relevant free energy density to be
minimized is the free energy per site with a fixed chemical potential $\mu $%
. Within our model, the experimental change in the strength in the periodic
potential corresponds to changing $\Delta /t_{b}Z.$

\begin{figure}[tbp]
\includegraphics[width=6.0cm,height=8.0cm]{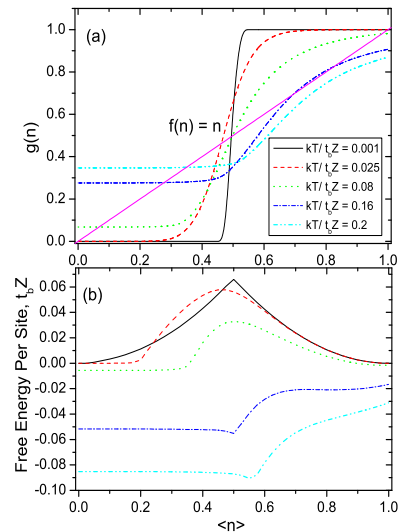}
\caption{Illustrations for $\Delta /t_{b}Z=B^{\ast }-2\times 10^{-7},$ where
$B^{\ast }$ is the zero-temperature transition point. \ (a) The
self-consistent criteria in which solutions are as the cross-over points of $%
g\left( \bar{n}\right) =\frac{\exp \left( -\protect\beta E_{1}\right) }{%
\frac{1}{2}+\exp \left( -\protect\beta E_{1}\right) }$ $=\bar{n}.$ \ As the
temperature arises, a middle branch of self-consistent $\bar{n}$ appears. \
The low and high $\bar{n}$ branches correspond respectively to the normal
solid and supersolid solutions. (b) The free energy per site as a function
of $\bar{n}$. \ The normal solid and supersolid solutions have free energy
around the two local minima. \ The first order SF-MI transition temperature
is $T_{c}/t_{b}Z\simeq 0.0246.$}
\label{nsolution}
\end{figure}

At zero temperature, there are no magnons. Using spin wave approximation,
the superfluid order parameter and the spin correlation are respectively $%
\left\langle b\right\rangle \simeq 0.478$ and $B^{\ast }=B\left( \bar{n}%
,T=0\right) \simeq 0.265$ for all non-zero $\bar{n}$'s. \ The
self-consistent Eq. (\ref{nEqu}) and its $T=0$ limit given by Eq. (\ref{free
energy}) indicate that when $\Delta /t_{b}Z<B^{\ast },$ $\bar{n}=1$ will be
the self-consistent solution that minimizes the total energy of the system,
but when $\Delta /t_{b}Z>B^{\ast },$ the ground state will have $\bar{n}=0.$
However, the $\bar{n}=1$ self-consistent solution branch does not disappear
and hence is metastable until $\Delta /t_{b}Z$ increases past $2B^{\ast }.$
\ Thus, there is a first order phase transition from the MI state to the SF
state with decreasing $\Delta /t_{b}Z.$ Accompanying the MI-SF transition is
a jump in boson density (or equivalently, vacancy density) and exciton
density . We thus recover the 'vacuum switching' first-order transition
obtained by DMZ\cite{DMZ} using the single-site MFT. \ Within single-site
MFT, the transition from MI to SF occurs at $\Delta /t_{b}Z=0.25,$ which is
less than our critical value of $\Delta /t_{b}Z=B^{\ast }$. This is the
consequence of the SF phase being stabilized against the MI phase due to
quantum phase fluctuations,

On the SF side, as the temperature rises from zero, thermal fluctuations
will decrease the SF order parameter $\left\langle b\right\rangle $ both
directly and also indirectly through the decrease in $\bar{n}.$ At the same
time, the decrease in spin correlation $B$ can drive the system from the
defect rich ($\bar{n}$ close to $1)$ to the defect poor ($\bar{n}$ close to $%
0)$ phase. As $T$ continues to increase, we can expect one of two scenarios.
The condensate amplitude may vanish while remaining in the defect rich
phase, followed subsequently by a transition from the defect rich into the
defect poor phase. \ Should that be the case, we expect $\left\langle
b\right\rangle $ to go to $0$ continuously. The other possibility is that
there is only one transition, which primarily is from defect rich to defect
poor, and when that happens, it drives the Bose condensed amplitude to $0.$
Our calculation shows that it is the second case that is realized provided $%
\Delta /t_{b}Z$ is not too far from $B^{*}$. This is shown in Fig \ref%
{nsolution}. \ The self-consistent solutions of $\bar{n}$ as $T$ is
increased are shown in Fig. \ref{nsolution}(a). We see that at finite $T,$
in between the $\bar{n}$ values of the solutions that evolve from the $\bar{n%
}=0$ and $\bar{n}=1$ solutions, there is a third self-consistent solution.
This solution is unstable, corresponding to a local maximum in the free
energy(Fig \ref{nsolution}(b)). \ The low and high $\bar{n}$ solutions are
local minima in the free energy. At $T=0,$ the high $\bar{n}$ solution is
the global minimum, but with increasing $T,$ the small $\bar{n}$ solution
becomes the free energy minimum\ at a critical temperature $T_{c}.$ Beyond, $%
T_{c},$ the large $\bar{n}$ solution becomes metastable, but it remains a
local minimum until a higher temperature where it merges with the with the
unstable middle branch solution. The above behavior is typical of first
order transition. In the present case, a jump in $\bar{n}$ occurs at $T_{c}.$
Just below $T_{c},$ $\bar{n}$ is large enough that the coupling $J$ in $%
H_{s} $ is sufficient for $\left\langle b\right\rangle $ to be non-zero.
Just above $T_{c,}$ the jump to a small $\bar{n}$ results in a value of $J$
such that $T_{c}$ is above the Bose condensation temperature, and hence $%
\left\langle b\right\rangle $ jumps to $0$ across the transition.

The value of $\Delta /t_{b}Z$ shown in Fig \ref{nsolution} is very close to $%
B^{\ast }$, yet $T_{c}$ is not that close to $0.$ The reason for this is
because of the presence of reentrance in the MI side of the transition. The
reentrance can be understood qualitatively as follows. Just on the MI side
of the transition, the energy of the defect poor state and the defect rich
state are almost degenerate. The defect poor phase has excitations with gap $%
\Delta ,$ and hence the difference of the free energy from the ground state
energy is exponentially small at low $T$. The defect rich state however is a
Bose condensate, and has gapless excitations. Thus, its free energy
decreases by a power of the temperature. As the temperature is raised, the
free energy gain can overcome the ground state energy difference between the
two phases, and the defect rich phase becomes more stable than the defect
poor phase. Thus as the temperature is raised, the system first undergos a
(first order) transition from the defect poor MI phase into the defect rich
SF phase, and then later on at a higher temperature undergoes the SF-MI
first order transition discussed previously back into the defect poor phase.

\begin{figure}[tbp]
\includegraphics[width=7.0cm,height=6.0cm]{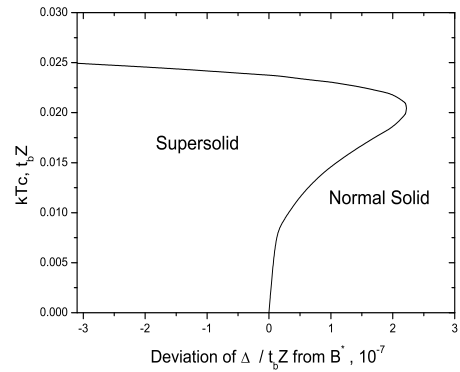}
\caption{The finite-temperature phase diagram, obtained by comparing free
energy per site, around the the zero temperature transition point $\Delta
/t_{b}Z=B^{\ast }\simeq 0.265.$ The tick labels of the horizontal axis
denote the deviation in the unit of $10^{-7}t_{b}Z$ from the zero
temperature transition point$.$ \ There is a region on the zero-temperature
normal solid side where the reentrance phenomenon occurs at the finite
temperature.\ }
\label{phase diagram per site}
\end{figure}

On a more quantitative level, the free energy of the defect poor phase at
low temperature is given by
\begin{equation*}
G_{0}\left( T\right) =-k_{B}T\left( \ln \left( 1+2\exp \left( -\Delta
/k_{B}T\right) \right) \right)
\end{equation*}%
On the other hand, for the defect rich phase, the dominant temperature
correction is due to phase excitations about the condensate, giving
\begin{equation*}
G_{1}\left( T\right) \approx \Delta -t_{b}ZB^{\ast }-\frac{\zeta \left(
3\right) }{\pi ^{2}\left( t_{b}\bar{n}^{2}\sqrt{Z}\right) ^{3}}\left(
k_{B}T\right) ^{4}
\end{equation*}%
where the Riemann zeta function $\zeta \left( 3\right) =$\ $1.202.$ If $%
G_{1}(T)<G_{0}(T),$ then the stable phase is the defect rich Bose condensed
phase. \ On the SF side ($\Delta <t_{b}ZB^{\ast }),$ $G_{1}(0)<G_{0}(0),$
and the difference is further enhanced at low temperature due to thermal
fluctuations. On the MI side, $G_{1}(0)>G_{0}(0),$ and the MI is the stable
ground state. But as $T$ increases, the $T^{4}$ correction in $G_{1}(T)$
dominate over the exponential correction in $G_{0}(T),$ and provided the
ground state energy difference is not too big, $G_{1}(T)$ becomes smaller
than $G_{0}(T)$ beyond some low temperature, giving rise to the 'reentrance'
phenomenon. As $T$ continues to increase, the disordering effect of thermal
fluctuations finally dominates so that there is a second transition back
into the defect poor MI phase. \ The temperature dependence of the free
energies of these two competing phases is illustrated in Fig. \ref{compare
free energy per site}. The phase diagram in the vicinity of the $T=0$ SF-MI
transition is shown in Fig. \ref{phase diagram per site}. The reentrance on
the MI side can be clearly seen.

\begin{figure}[tbp]
\includegraphics[width=6.0cm,height=7.0cm]{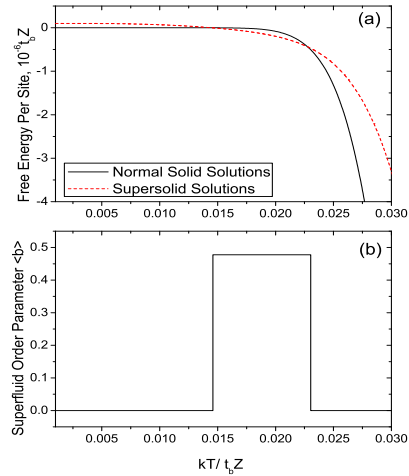}
\caption{Illustrations for $\Delta /t_{b}Z=B^{\ast }+10^{-7},$ where $%
B^{\ast }$ is the zero-temperature transition point. \ (a) The free energy
per site plots of the normal solid and the supersolid self-consistent
solutions as functions of temperature. \ There are two phase transitions,
both first-ordered, as the temperature increases for this specific $\Delta .$
\ (b) \ The solved superfluid order parameter $<b>$ as a function of
temperature, showing sudden jumps at the two first-ordered phase transition
points. \ }
\label{compare free energy per site}
\end{figure}

\begin{figure}[tbp]
\includegraphics[width=7.0cm,height=6.0cm]{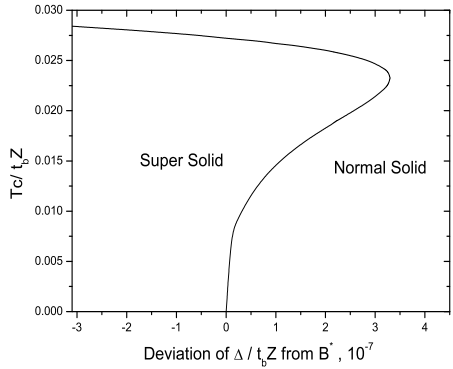}
\caption{The finite-temperature phase diagram around the the zero
temperature transition point $\Delta /t_{b}Z=B^{\ast }\simeq 0.265$,
obtained by comparing free energy per particle rather than the value per
site as shown in Fig. \protect\ref{phase diagram per site}. $\ $The result
compared to Fig. \protect\ref{phase diagram per site} shows moderate
difference. \ }
\label{phase diagram per atom}
\end{figure}

Although our results are based on our modified spin wave theory for the spin
Hamiltonian $H_{s},$ we believe the results are quite reliable. As a
comparison, we consider the case of $1D,$ where $H_{s}$ can be solved
exactly using the well-known Jordan-Wigner transformation. The $1D$ case by
itself is also interesting because it can be realized experimentally in cold
atoms. Using the Jordan-Wigner transformation, \ the spin 1/2 operators are
mapped into spinless fermion operators
\begin{eqnarray}
g_{l} &=&e^{i\phi _{l}}S_{l}^{-}  \notag \\
g_{l}^{+} &=&e^{-i\phi _{l}}S_{l}^{+} \\
g_{l}^{+}g_{l} &=&S_{l}^{+}S_{l}^{-}=\frac{1}{2}+S_{l}^{z}  \notag
\end{eqnarray}
where
\begin{equation*}
\phi _{l}=\pi \sum_{i=1}^{l-1}g_{i}^{+}g_{i}
\end{equation*}
and $\left\{ g_{l},g_{m}^{+}\right\} =\delta _{lm},\left\{
g_{l},g_{m}\right\} =0.$ Then the spin Hamiltonian can be rewritten by the
fermion $g$ operators and be diagonalized in $k$ space as
\begin{equation}
H_{s}=-t_{b}\bar{n}^{2}\sum_{i}\left( \hat{S}_{i}^{+}\hat{S}_{i+1}^{-}+\hat{S%
}_{i}^{+}\hat{S}_{i-1}^{-}\right) =\sum_{k}\varepsilon _{k}g_{k}^{+}g_{k}
\end{equation}
with
\begin{equation}
\varepsilon _{k}=-t_{b}Z\bar{n}^{2}\cos k
\end{equation}
The spin correlation is then found as
\begin{equation}
B=\frac{1}{N}\sum_{k}\cos k\left\langle g_{k}^{+}g_{k}\right\rangle
\end{equation}
where
\begin{equation}
\left\langle g_{k}^{+}g_{k}\right\rangle =\left( \exp \left( \beta
\varepsilon _{k}\right) +1\right) ^{-1}
\end{equation}
is the fermion distribution function. \ At $T=0,$ the negative energy levels
are filled, while the positive ones are empty . \ The zero temperature
transition is at $\Delta /t_{b}Z=B^{**}=\frac{1}{\pi }\approx 0.318.$ \ We
note that in addition to solving $H_{s}$ exactly, this result is actually
the exact result for the Hamiltonian Eq. (\ref{spinxy}). This is because $%
\left[ H,\widehat{n}_{i}\right] $\ $=0$ , so $n_{i}$ is a good quantum
number for the ground state. Assuming no breaking of translational
invariance, we then have either all $n_{i}=0$ or all $n_{i}=1.$ Comparing
the exact solution to the modified spin-wave method, which gives $\Delta
/t_{b}Z=B_{1D}^{*}=0.314$ as the transition point in 1D, we see that the
approximate modified spin wave method does very well. \ We also note that
both methods show that there is no Bose condensation in $1D$ at $T=0,$ and
the transition in $\overline{n}$ is driven by short range spin correlation $%
B $.\

\begin{figure}[tbp]
\includegraphics[width=9.0cm,height=8.0cm]{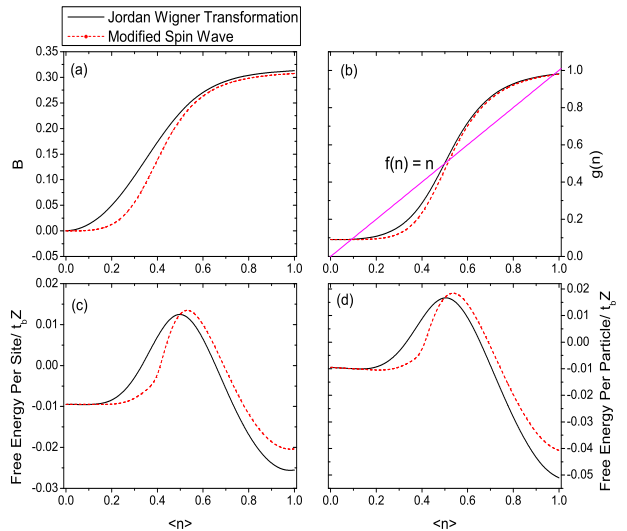}
\caption{1D MFT results of (a) correlation function $B\left( \bar{n}%
,T\right) ,$ (b) the self-consistent scheme in finding $\bar{n}$,
(c) free energy per site and (d) free energy per particle by
Jordan-Wigner
transformation and the modified spin wave method. \ System parameters are $%
\Delta /t_{b}Z=0.3$ and $kT/t_{b}Z=0.1.$} \label{1Djordan}
\end{figure}

\begin{figure}[tbp]
\includegraphics[width=6.0cm,height=8.0cm]{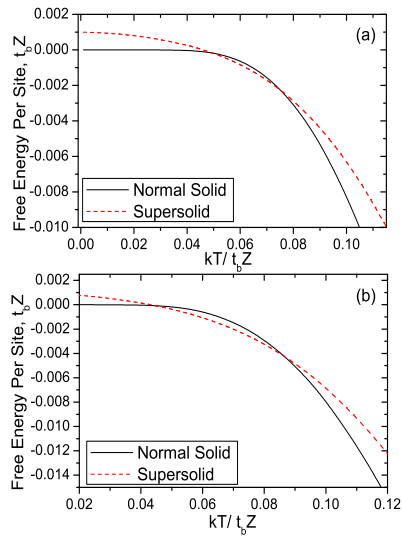}
\caption{1D MFT results of free energy per site of high $\bar{n}$ and low $%
\bar{n}$ self-consistent solutions indicating phase transitions by
(a) modified spin wave approximation at $\Delta
/t_{b}Z=B_{1D}^{\ast }+0.001$ and by (b) Jordan-Wigner
transformation at\ $\Delta /t_{b}Z=B^{\ast \ast }+0.001.$}
\label{jordan free energy per site}
\end{figure}

At finite $T,$ the free energy due to $H_{s}$ can still be calculated
exactly in $1D,$ but the MFT procedure we use to decouple $\widehat{S}$ and $%
\widehat{n}$ is an approximation. The MF finite temperature
results for the 1D case by both the modified spin wave and
Jordan-Wigner transform are shown in Fig. \ref{1Djordan} and
\ref{jordan free energy per site}. \ The two method gives similar
predictions in the spin correlation and the self-consistent $n$
solutions. \ The results of free energy obtained by the two
methods exhibit some difference, but the lower the temperature the
smaller the discrepancy. \ More importantly, the reentrance
behavior from defect poor to defect rich back to defect poor
phases on the MI side appears in both methods. \ Both methods give
phase transitions in $\bar{n}$ at finite temperatures in 1D, which
is of course incorrect and an artifact of MFT, and the transitions
should be interpreted as crossovers in defect densities rather
than real phase transitions. Nevertheless, the $1D$ results
provide additional support for the 3D phase diagram presented
above.


The results presented so far are relevant for bosons in optical lattice,
where the experiments may be performed at constant chemical potential and
the lattice constant is fixed by the optical lattice. To apply them to
supersolid $^{4}$He, modifications are necessary for the following reason.
In the case of solid $^{4}$He, the lattice constant is not fixed externally
but self-determined. Instead it is the number of helium atoms that is fixed.
Therefore, when comparing the free energy of the defect poor and defect rich
phase, one should use the free energy per atom rather than the free energy
per site, with $N=N_{a}+N_{b}+N_{v},$ where $N$ is the number of sites, $%
N_{a,b}$ are the number of sites occupied by helium atoms in $a$ and $b$
states, and $N_{v}$ the number of vacancies. \ The number of $^{4}$He atoms $%
N_{He}=N_{a}+N_{b}.$ Minimizing the mean field free energy per atom instead
of per site gives small quantitative corrections, but do not change the main
results presented above. The phase diagram in that case is shown in Fig. \ref%
{phase diagram per atom}. The transition temperature is changed, but the
first order transitions and reentrance remain.

In summary, we have shown that the MI-SF transition of a two-band boson
Hubbard model can differ significantly from that of the one-band model.
Instead of a continuous transition, the transition here is first order, both
for the $T=0$ and the finite $T$ transitions. On the MI side, there can be a
reentrant transition into the SF phase and then back into the MI phase as
temperature is raised.\ \ The underlying physics is that the MI-SF
transition is actually a by-product of the transition between two competing
phases, which are the defect poor and defect rich phases. In the defect poor
phase, bosons are localized, while in the defect rich phase, they are
delocalized and Bose condense. The defect rich phase contain a finite
density of excitons (defined in the Introduction section) and vacancies even
at $T=0.$ Thus, the MI-SF transition is accompanied by a change in boson
density in the case of optical lattice and a change from commensurate to
incommensurate solid in the case of solid $^{4}$He. Our results confirm that
the results obtained by DMZ are robust when phase fluctuations are included,
and also extend their $T=0$ results to finite temperature. For $3D,$ these
fluctuations are taken into account using a modified spin wave theory.
Although it is an approximate method, we believe it to be quantitatively
reliable for our model by comparing the results in $1D$ using this method
and using Jordan-Wigner transformation.

This work was partly supported by the Hong Kong's RGC grant 706206.


\begin{thebibliography}{99}
\bibitem{Fisher 1989} M. P. A. Fisher, P. b. Weichman, G. Grinstein, D. S.
Fisher, Phys. Rev. B \textbf{40}, 546 (1989)

\bibitem{Kim Chan Nature} E. Kim and M. H. W. Chan, Nature \textbf{427}, 225
(2004)

\bibitem{Kim Chan Science} E. Kim and M. H. W. Chan, Science \textbf{305},
1941 (2004)

\bibitem{Bloch Nature} M. Greiner, O. Mandel, T. Esslinger, T. W. Hansch, I.
Bloch, Nature \textbf{415}, 6867 (2002)

\bibitem{DMZ} X. Dai, M. Ma, F. C. Zhang, Phys. Rev. B \textbf{72}, 132504
(2005)

\bibitem{Andreev Lifshitz 1969} A. F. Andreev and I. M. Lifshitz, Soviet
Phys. JETP \textbf{29}, 1107 (1969)

\bibitem{Chester 1970} G. V. Chester, Phys. Rev. A \textbf{2}, 256 (1970)

\bibitem{Leggett 1970} A. J. Leggett, Phys. Rev. Lett. \textbf{25}, 1543 -
1546 (1970)

\bibitem{Rittner Reppy 2006} A. S. C. Rittner, J. D. Reppy, Phys. Rev. Lett.
\textbf{97}, 165301 (2006)

\bibitem{Shirahama 2006} M. Kondo, S. Takada, Y. Shibayama, K. Shirahama,
cond-mat/0607032(2006)

\bibitem{Kubota 2007} A. Penzev, Y. Yasuta, M. Kubota, cond-mat/0702632
(2007)

\bibitem{Burovsky 2005} E. Burovski, E. Kozik, A. Kuklov, N. Prokof'ev, B.
Svistunov, Phys. Rev. Lett. \textbf{94}, 165301 (2005)

\bibitem{Day Beamish 2006} J. Day, J. Beamish, Phys. Rev. Lett. \textbf{96}%
,105304 (2006)

\bibitem{Sasaki 2006} S. Sasaki, R. Ishiguro, F. Caupin, H. J. Maris, S.
Balibar, Sience \textbf{313}, 1098 (2006)

\bibitem{Anderson Brinkman Huse 2005} P. W. Anderson, W. F. Brinkman, D. A.
Huse, Science \textbf{310} 1164 (2005)

\bibitem{Simmons 1989} B. A. Frasass, P. R. Granfors, and R. O. Simmons,
Phys. Rev. B \textbf{39}, 124 (1989)

\bibitem{Clark Ceperley 2006} B. K. Clark, D. M. Ceperley, Phys. Rev. Lett.
\textbf{96}, 105302 (2006)

\bibitem{Nosanow 1962} L. H. Nosanow, G. L. Shaw, Phys. Rev. \textbf{128},
546 - 550 (1962)

\bibitem{Jaksch 1998} D. Jaksch, C. Bruder, J. I. Cirac, C. W. Gardiner, P.
Zoller, Phys. Rev. Lett. \textbf{81}, 3108 - 3111 (1998)

\bibitem{Jaksch 2004} S. R. Clark, D. Jaksch, Phys. Rev. A \textbf{70},
043612 (2004)

\bibitem{theja} T. N. De Silva, M. Ma, F. C. Zhang, Phys. Rev. B \textbf{66}%
, 104417 (2002)

\bibitem{Gomez-Santos Joannopoulos 1987} G. Gomez-Santos and J. D.
Joannopoulos, Phys. Rev. B \textbf{36}, 8707 (1987)

\bibitem{Takahashi 1987} M. Takahashi, Phys. Rev. Lett \textbf{58}, 168,
(1987)

\bibitem{Takahashi 1989} M. Takahashi, Phys. Rev. B \textbf{40}, 2494 (1989)

\bibitem{Hirsch Tang 1989} J. E. Hirsch and S. Tang, Phys. Rev. B \textbf{40}%
, 4769 (1989)

\bibitem{footnote} The finite temperature transition can be second order in
the very deep SF phase far from the zero temperature transition point. \
However, as such transition, led by the annihilation of the long range
order, occurs only when the defect free state, the vacancy and the exciton
are all nearly degenerate, it is out of our concern investigating the
two-band model in this work. \
\end{thebibliography}
\end{document}